\documentclass[12pt]{article} 
\usepackage{epsfig} 
\def\lsim{\mathrel{\rlap{\lower4pt\hbox{\hskip1pt$\sim$}}
    \raise1pt\hbox{$<$}}}                
\def\gsim{\mathrel{\rlap{\lower4pt\hbox{\hskip1pt$\sim$}}
    \raise1pt\hbox{$>$}}}                
\newcommand{\zp}[3]{Z.\ Phys.\ {\bf C#1} (19#2) #3}
\newcommand{\pl}[3]{Phys.\ Lett.\ {\bf B#1} (19#2) #3}
\newcommand{\plold}[3]{Phys.\ Lett.\ {\bf #1B} (19#2) #3}
\newcommand{\np}[3]{Nucl.\ Phys.\ {\bf B#1} (19#2) #3}
\newcommand{\prd}[3]{Phys.\ Rev.\ {\bf D#1} (19#2) #3}
\newcommand{\prl}[3]{Phys.\ Rev.\ Lett.\ {\bf #1} (19#2) #3}

\newcommand{\mpl}[3]{Mod.\ Phys.\ Lett.\ {\bf A#1} (19#2) #3}

\newcommand{\md}{\mbox{d}}
\def\simgt{\rlap{\lower 3.5 pt \hbox{$\mathchar \sim$}} \raise 1pt \hbox {$>$}}
\def\simlt{\rlap{\lower 3.5 pt \hbox{$\mathchar \sim$}} \raise 1pt \hbox {$<$}}

\newcommand{\beq}{\begin{equation}}
\newcommand{\eeq}{\end{equation}}
\newcommand{\bea}{\begin{eqnarray}}
\newcommand{\eea}{\end{eqnarray}}

\newcommand{\jpsi}{{J\!/\!\psi}}
\newcommand{\as}{\mbox{$\alpha_s$}}
\newcommand{\beqa}{\begin{eqnarray}}
\newcommand{\eeqa}{\end{eqnarray}}
\begin{document}
\vspace*{1cm}
\begin{center}  \begin{Large} \begin{bf}
Prospects for Quarkonium Physics at HERA\\
  \end{bf}  \end{Large}
  \vspace*{5mm}
  \begin{large}
Matteo Cacciari and Michael Kr\"amer\\ 
  \end{large}

\vspace*{2mm}

Deutsches~Elektronen-Synchrotron~DESY,~D-22607~Hamburg,~FRG\\
\end{center}
\begin{quotation}
\noindent
{\bf Abstract:}
We work out and review prospects for future quarkonium physics at
\mbox{HERA}. We focus on the impact of color-octet contributions and
discuss how measurements at \mbox{HERA} can be used to test the
picture of quarkonium production as developed in the context of the
NRQCD factorization approach.
\end{quotation}
%
\section{Introduction}

The production of heavy quarkonium states in high energy collisions
provides an important tool to study the interplay between perturbative
and nonperturbative QCD dynamics. Quarkonium production in deep
inelastic scattering and photon-proton collisions at HERA has been
analysed at some length in the context of the previous HERA workshops
(see e.g.\ ref.\cite{JST92}). However, most of the previous studies
have been carried out within the color-singlet model (CSM) \cite{CS}
or the color-evaporation model \cite{CEM}. Only recently, a rigorous
theoretical framework for treating quarkonium production and decays
has been developed in ref.\cite{BBL95}.  The factorization approach is
based on the use of non-relativistic QCD (NRQCD) \cite{CLP} to
separate the short-distance parts from the long-distance matrix
elements and explicitly takes into account the complete structure of
the quarkonium Fock space.  This formalism implies that so-called
color-octet processes, in which the heavy-quark antiquark pair is
produced at short distances in a color-octet state and subsequently
evolves nonperturbatively into a physical quarkonium, should
contribute to the cross section.  According to the factorization
formalism, the inclusive cross section for the production of a
quarkonium state $H$ can be expressed as a sum of terms, each of which
factors into a short-distance coefficient and a long-distance matrix
element:
\beq\label{eq_fac}
\md\sigma(ep \to H + X) 
= \sum_n \md\hat{\sigma}(ep \to Q\overline{Q}\, [n] + X)\, 
         \langle {\cal{O}}^{H}\,[n]\rangle \, .
\label{eq:bbl}
\eeq
Here, $\md\hat{\sigma}$ denotes the short-distance cross section for
producing an on-shell $Q\overline{Q}$-pair in a color, spin and
angular-momentum state labelled by $n$.  The NRQCD vev matrix elements
$\langle {\cal{O}}^{H} \, [n] \rangle$ give the
probability for a $Q\overline{Q}$-pair in the state $n$ to form the
quarkonium state $H$. The relative importance of the various terms
in (\ref{eq_fac}) can be estimated by using NRQCD velocity scaling
rules \cite{LMNMH92}. For $v\to 0$ ($v$ being the average velocity of
the heavy quark in the quarkonium rest frame) each of the NRQCD
matrix elements scales with a definite power of $v$ and the general
expression (\ref{eq_fac}) can be organized into an expansion in powers
of $v^2$.

It has recently been argued in refs.\cite{TEV1,CL96} that quarkonium
production in hadronic collisions at the Tevatron can be accounted for
by including color-octet processes and by adjusting the unknown
long-distance color-octet matrix elements to fit the data. In order to
establish the phenomenological significance of the color-octet
mechanism it is however necessary to identify color-octet
contributions in different production processes.\footnote{Quarkonium
  production via color-octet states has also been studied in $e^+e^-$
  annihilation and $Z$ decays, for hadronic collisions in the energy
  range of fixed-target experiments and in $B$ decays \cite{BC95}.} In
this report we will focus our discussion on the prospects of
extracting information on the color-octet processes from the
measurements of inelastic quarkonium production at \mbox{HERA}.
Elastic/diffractive mechanisms will be discussed elsewhere
\cite{elastic}. We shall briefly review the impact of color-octet
contributions and higher-order QCD corrections on the cross section
for $\jpsi$ photoproduction and compare the theoretical predictions
with recent experimental data.  In this context we will also comment
on the possibility of measuring the gluon distribution in the proton
from $\jpsi$ photoproduction.  The high-statistics data to be
expected in the future at HERA will allow for a detailed comparison of
the theoretical predictions with experimental data not only for
$\jpsi$ photoproduction, but also for various other channels and
final states, like photoproduction of $\psi'$, $\Upsilon$ and $\chi$
states, associated $\jpsi + \gamma$ production, fragmentation and
resolved-photon contributions as well as deep inelastic $\jpsi$
production. We will discuss how these reactions can be used to
constrain the color-octet matrix elements and test the picture of
quarkonium production as developed in the context of the NRQCD
factorization approach.

\section{$\jpsi$ photoproduction}
Quarkonium production in high energy $ep$ collisions at \mbox{HERA} is
dominated by photoproduction events where the electron is scattered by
a small angle producing photons of almost zero virtuality. The
measurements at \mbox{HERA} provide information on the dynamics of
quarkonium photoproduction in a wide kinematical region,
$30~\mbox{GeV} \; \simlt \; \sqrt{s\hphantom{tk}}\!\!\!\!\!  _{\gamma
  p}\;\simlt\; 200~\mbox{GeV}$, corresponding to initial photon
energies in a fixed-target experiment of $450~\mbox{GeV} \; \simlt \;
E_\gamma \; \simlt \;$ $20,000~\mbox{GeV}$.  The production of $\jpsi$
particles in photon-proton collisions proceeds predominantly through
photon-gluon fusion. Elastic/diffractive mechanisms can be eliminated
by measuring the $\jpsi$ energy spectrum, described by the scaling
variable $z = {p\cdot k_\psi}\, / \, {p\cdot k_\gamma}$, with $p,
k_{\psi,\gamma}$ being the momenta of the proton and $\jpsi$,
$\gamma$ particles, respectively. In the proton rest frame, $z$ is the
ratio of the $\jpsi$ to $\gamma$ energy, $z=E_{\psi}/E_\gamma$. For
elastic/diffractive events $z$ is close to one; a clean sample of
inelastic events can be obtained in the range $z\;\simlt\;0.9$.

For $\jpsi$ production and at leading order in $v^2$, the general
factorization formula (\ref{eq_fac}) reduces to the standard
expression of the color-singlet model \cite{CS}. The short-distance
cross section is given by the subprocess
\begin{equation}\label{eq_cs}
\gamma + g \to c\bar{c}\, [^3S_1,\underline{1}] + g
\end{equation}
with $c\bar{c}$ in a color-singlet state (denoted by
\mbox{$\underline{1}$}), zero relative velocity, and
spin/angular-mo\-men\-tum quantum numbers $^{2S+1}L_J = {}^3S_1$.
Relativistic corrections due to the motion of the charm quarks in the
$\jpsi$ bound state enhance the large-$z$ region, but can be
neglected in the inelastic domain \cite{REL}. The calculation of the
higher-order perturbative QCD corrections to the short-distance
process (\ref{eq_cs}) has been performed in refs.\cite{KZSZ94,MK95}.
Inclusion of the NLO corrections reduces the scale dependence of the
theoretical prediction and increases the color-singlet cross section
by more than 50\%, depending in detail on the photon-proton energy and
the choice of parameters \cite{MK95}.
 
Color-octet configurations are produced at leading order in
$\mbox{$\alpha_{\mbox{\scriptsize s}}$}$ through the $2\to 1$ parton
processes \cite{CK96,AFM,KLS}
\begin{eqnarray}\label{eq_oc0}
\gamma + g &\! \to \!& c\bar{c}\, [^1S_{0},\underline{8}]
\nonumber \\
\gamma + g &\! \to \!& c\bar{c}\, [^3P_{0,2},\underline{8}]
\,.
\end{eqnarray}
Due to kinematical constraints, the leading color-octet terms will
only contribute to the upper endpoint of the $\jpsi$ energy spectrum,
$z\approx 1$ and $p_T\approx 0$, $p_T$ being the $\jpsi$
transverse momentum. It has, however, been argued in 
ref.\cite{BFY} that sizable higher-twist effects are expected to
contribute in the region $p_T\; \simlt\; 1$~GeV, which cause the
breakdown of the factorization formula (\ref{eq_fac}).  Moreover,
diffractive production mechanisms which cannot be calculated within
perturbative QCD might contaminate the region $z\approx 1$ and make it
difficult to extract precise information on the color-octet processes.

\begin{figure}[t]
\vspace*{3cm}
\hspace*{-10mm}
\begin{picture}(7,7)
\includegraphics{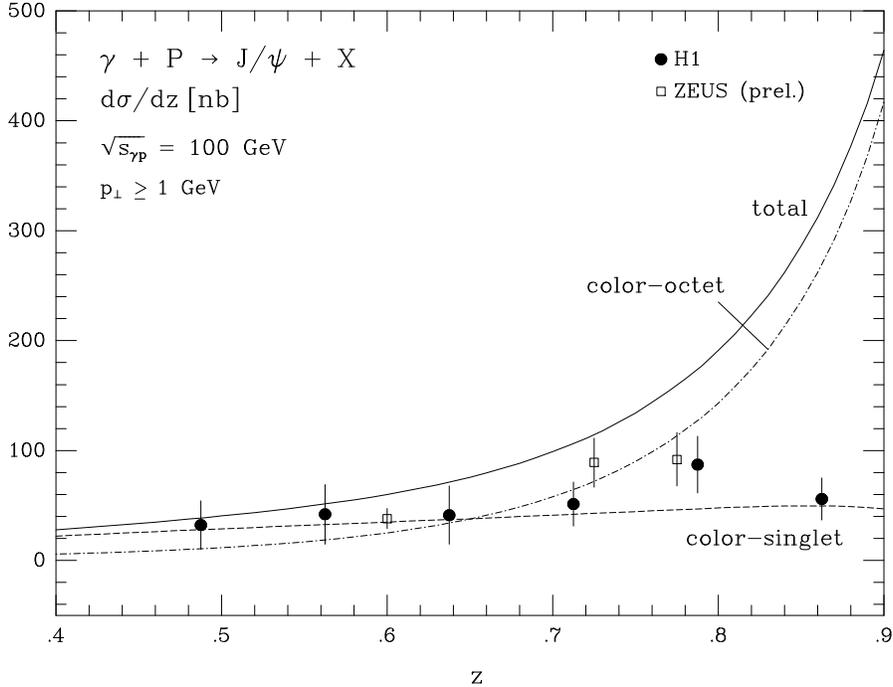}
\end{picture}
\vspace*{6.cm}
\begin{center}
\parbox{12cm}{
\caption[dummy]{\label{fig_3} \small
  Color-singlet and color-octet contributions to 
  the $J\!/\!\psi$ energy distribution $\md\sigma/\md{}z$ at the
  photon-proton centre of mass energy $\sqrt{s\hphantom{tk}}\!\!\!\!\!
  _{\gamma p}\,\, = 100$~GeV integrated in the range $p_T \ge
  1$~GeV \cite{CK96}.  Experimental data from \cite{H1,ZEUS}.}
}
\end{center}
\end{figure}

It is therefore more appropriate to study $\jpsi$ photoproduction in
the inelastic region $z \le 0.9$ and $p_T \ge 1$~GeV where no
diffractive channels contribute and where the general factorization
formula (\ref{eq_fac}) and perturbative QCD calculations should be
applicable.  Color-octet configurations which contribute to inelastic
$\jpsi$ photoproduction are produced through the subprocesses
\cite{CK96,KLS}
\begin{eqnarray}\label{eq_oc2}
\gamma + g &\! \to \!& c\bar{c}\, [^1S_{0},\underline{8}] 
  + g \nonumber \\
\gamma + g &\! \to \!& c\bar{c}\, [^3S_{1},\underline{8}] 
  + g \nonumber \\
\gamma + g &\! \to \!& c\bar{c}\, [^3P_{0,1,2},\underline{8}] + g 
\,.
\label{jpsioctet}
\end{eqnarray}
Light-quark initiated contributions are strongly suppressed at
\mbox{HERA} energies and can safely be neglected.  Adopting NRQCD
matrix elements consistent with those extracted from the fits to
prompt $\jpsi$ data at the Tevatron \cite{CL96} (see Table
\ref{table1}) one finds that color-octet and color-singlet
\begin{table}[htb]
\begin{center}
\begin{tabular}{|c|c|c|} 
\hline
$\langle {\cal{O}}^{J\!/\!\psi}\,[^{3}S_{1},\underline{1}]\rangle 
\hphantom{/m_c^2}$ &$1.16$ GeV$^3$ & $\quad m_c^3 v^3$ \\ 
$\langle {\cal{O}}^{J\!/\!\psi}\,[^{1}S_{0},\underline{8}]\rangle
\hphantom{/m_c^2}$ &$10^{-2}$ GeV$^3$ & $\quad m_c^3 v^7$ \\ 
$\langle {\cal{O}}^{J\!/\!\psi}\,[^{3}S_{1},\underline{8}]\rangle
\hphantom{/m_c^2}$ &$10^{-2}$ GeV$^3$ & $\quad m_c^3 v^7$ \\ 
$\langle {\cal{O}}^{J\!/\!\psi}\,[{}^{3}P_{0},\underline{8}]\rangle / m_c^2$  
& $10^{-2}$ GeV$^3$ & $\quad m_c^3 v^7$ \\
\hline
\end{tabular}
\parbox{12cm}{
\caption{
\label{table1}
\small Values of the NRQCD matrix elements used in the numerical
 analysis, with the velocity and mass scaling. $v$ is the velocity of
 the heavy quark in the quarkonium rest frame. For charmonium it holds
 $v^2 \simeq 0.25$.} }
\end{center}
\end{table}
contributions to the inelastic cross section are predicted to be of
comparable size \cite{CK96,KLS}.  However, taking into account the
uncertainty due to the value of the charm quark mass and the strong
coupling, the significance of color-octet contributions can not easily
be deduced from the analysis of the absolute $\jpsi$ production rates.
A distinctive signal for color-octet processes should, however, be
visible in the $\jpsi$ energy distribution
$\mbox{d}\sigma/\mbox{d}{}z$ shown in fig.\ref{fig_3} \cite{CL96}.
Since the shape of the distribution is insensitive to higher-order QCD
corrections or to the uncertainty induced by the choice for $m_c$ and
$\mbox{$\alpha_{\mbox{\scriptsize s}}$}$, the analysis of the $\jpsi$
energy spectrum $\mbox{d}\sigma/\mbox{d}{}z$ should provide a clean
test for the underlying production mechanism.  From fig.\ref{fig_3} we
can conclude that the shape predicted by the color-octet contributions
is not supported by the experimental data and that the $\jpsi$ energy
spectrum is adequately accounted for by the color-singlet channel. This is
however not to be considered as a failure of the factorization approach, since
also other analyses (see for instance the last item of ref. \cite{BC95})
have pointed out that the fits to the Tevatron data may have returned
slightly too large values for the matrix elements.
With higher statistics data it will be possible to extract more
detailed information on the color-octet matrix elements, in particular
from the analysis of the $J\!/\!\psi$ energy distribution in the
inelastic region.

\begin{figure}[t]
\vspace*{3cm}
\hspace*{-10mm}
\begin{picture}(7,7)
\includegraphics{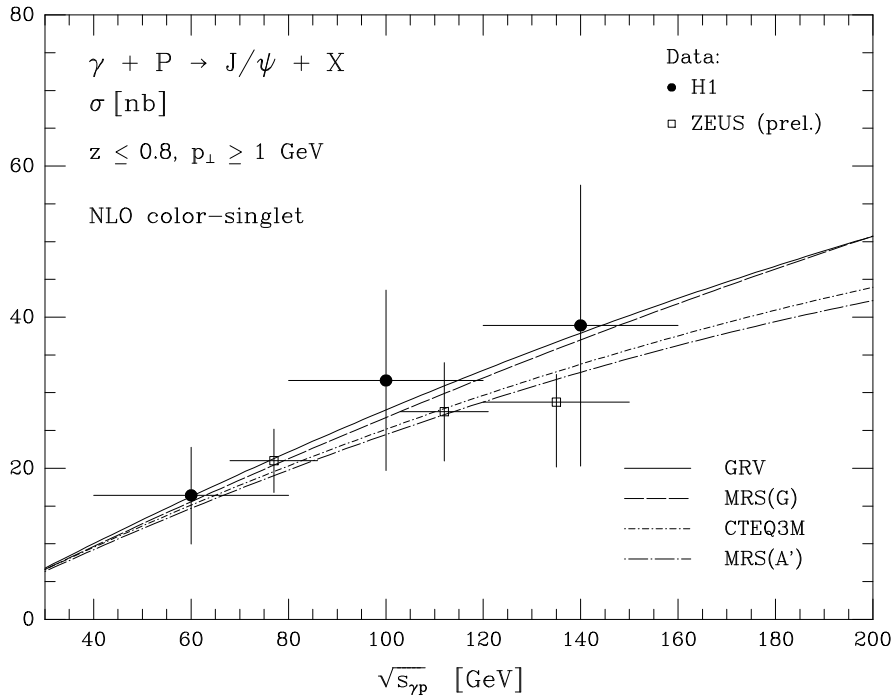}
\end{picture}
\vspace*{6cm}
\begin{center}
\parbox{12cm}{
\caption[dummy]{\label{fig_5} \small
  NLO color-singlet prediction for the total 
  inelastic $J\!/\!\psi$ photoproduction cross section as a function
  of the photon-proton energy for different parametrizations \cite{pdfs} 
  of the parton distribution in the proton \cite{MK95}. 
  Experimental data from \cite{H1,ZEUS}.}
}
\end{center}
\end{figure}

The impact of higher-order QCD corrections on total cross sections and
differential distributions has been studied thoroughly for the
color-singlet channel in ref.\cite{MK95}. A detailed analysis of the
spectra in the high energy range at \mbox{HERA} shows that the
perturbative calculation is not well-behaved in the limit $p_T \to
0$, where $p_T$ is the transverse momentum of the $\jpsi$.  No
reliable prediction can be made in this singular boundary region
without resummation of large logarithmic corrections caused by
multiple gluon emission. If the small $p_T$ region is excluded
from the analysis, experimental results on differential distributions
and total cross sections are well accounted for by the color-singlet
channel alone, including NLO QCD corrections, see e.g.\ 
fig.\ref{fig_5}.  However, since the average momentum fraction of the
partons is shifted to larger values when excluding the small-$p_T$
region, the sensitivity of the prediction to the small-$x$ behaviour
of the gluon distribution is not very distinctive.  A detailed
analysis reveals that the size of the QCD corrections increases when
adopting parton densities with flatter gluons.  The sensitivity to
different gluon distributions is thus reduced in next-to-leading order
as compared to the leading-order result, in particular when choosing a
small charm mass and a large value for the strong coupling.
Parametrizations with extremely flat gluons like MRS(D0') \cite{MRSD0}
are clearly disfavoured by the recent \mbox{HERA} measurements of the
proton structure function \cite{HERAF2} and do not allow for a
reliable prediction in the high energy region.  For the parameters
adopted in fig.\ref{fig_5}, the MRS(D0') distribution leads to
next-to-leading order results not very different from those obtained
with the MRS(A') parametrization.  The corresponding $K$-factors are
however uncomfortably large, $K \sim 4$, casting doubts on the
reliability of the perturbative expansion as obtained by using flat
gluon distributions. If parton distributions with steep gluon
densities are adopted, the next-to-leading order cross section is
well-behaved and gives an adequate description of the experimental
data, as demonstrated in fig.\ref{fig_5}.

The cross section for the production of $\psi'$ particles has been
measured by several photo- and hadroproduction experiments
\cite{psiidata} to be suppressed by a factor $\sim 0.25$ compared to
$\jpsi$ production.  This result is consistent with naive estimates
obtained within the color-singlet model where one assumes that the
effective charm masses in the short distance amplitudes scale like the
corresponding $\psi'$ and $\jpsi$ masses \cite{CS}. Within the NRQCD
factorization approach it is however conceptually preferred to express
the short-distance coefficients in terms of the charm quark mass
rather than the quarkonium mass, yielding $\sigma(\psi')/\sigma(\jpsi)
\sim 0.5$ for color-singlet dominated production channels.
Relativistic corrections and color-octet contributions are expected to
affect the ratio $\sigma(\psi')/\sigma(\jpsi)$, but no quantitative
prediction can be made with the present experimental and theoretical
information.

Bottomonium production is a particular interesting subject to be
studied at HERA. The larger value of the bottom quark mass makes the 
perturbative QCD predictions of the short-distance cross section more
reliable than for charm production. Moreover, the derivation of the
factorization formula eq.(\ref{eq_fac}) relies on the fact that the
momentum scales which govern the bound state dynamics are well
separated: $(m_Q v^2)^2 \ll (m_Q v)^2 \ll m_Q^2$. This assumption is
reasonably good for charmonium (where $v^2\sim 0.3$) but very good for
bottomonium (where $v^2\sim 0.1$). Thus, the theoretical predictions
for bottomonium production should be much more reliable than those for
charmonium. The production rates for $\Upsilon$ bound states are,
however, suppressed, compared with $\jpsi$ states, by a factor of
about 300 at \mbox{HERA}, a consequence of the smaller bottom electric
charge and the phase space reduction by the large $b$ mass. 

\section{$\jpsi$ production via fragmentation}
At sufficiently large transverse momentum $p_T$, quarkonia
production is dominated by fragmentation, the production of a parton
with large $p_T$ which subsequently decays into the quarkonium
state and other partons \cite{BY93}. While the fragmentation
contributions are of higher order in $\alpha_s$ compared to direct
quarkonia production in fusion processes, they are enhanced by powers
$p_T^2/m_c^2$ and can thus overtake the fusion contribution at
$p_T \gg m_c$. It has in fact been argued that quarkonium production at
large $p_T$ in hadronic collisions at the Tevatron can be
accounted for by including gluon fragmentation into color-octet states
\cite{TEV1}.

The fragmentation contribution to the differential cross section for
producing a quarkonium state $H$ at large $p_T$ can be written in
the factorized form
\begin{equation}
\mbox{d}\sigma(\gamma+p\to H+X)=\sum_i\int_0^1\mbox{d}\zeta\, 
\mbox{d}\hat{\sigma}(\gamma+p\to i+X)\,D_{i\to H}(\zeta)
\end{equation}
where $\mbox{d}\hat{\sigma}$ is the differential cross section for
producing a parton of type $i$ with momentum $p_T/\zeta$. The
fragmentation function $D_{i\to H}$ gives the probability that a jet
initiated by parton $i$ contains a hadron $H$ carrying a fraction
$\zeta$ of the parton momentum. According to the NRQCD factorization
formalism, fragmentation functions for charmonium have the general
form
\begin{equation}
D_{i\to H}(\zeta,\mu) = \sum_n d_{i\to c\bar{c}[n]X}(\zeta,\mu)\, 
  \langle {\cal{O}}^{H}\,[n]\rangle 
\end{equation}
analogous to eq.(\ref{eq_fac}). The function $d_{i\to c\bar{c}[n]X}$
gives the probability for the parton $i$ to form a jet that includes a
$c\bar{c}$ pair in the state labelled by $n$. It can be calculated at
an initial scale $\mu\sim m_c$ as a perturbative expansion in
$\alpha_s(m_c)$ \cite{BY93,BY94} and be evolved up to higher scales
$\mu\sim p_T$ by using the Altarelli-Parisi evolution equations.

\begin{figure}[tb]

\vspace*{-5mm}
\hspace*{1.5cm}
\epsfig{file=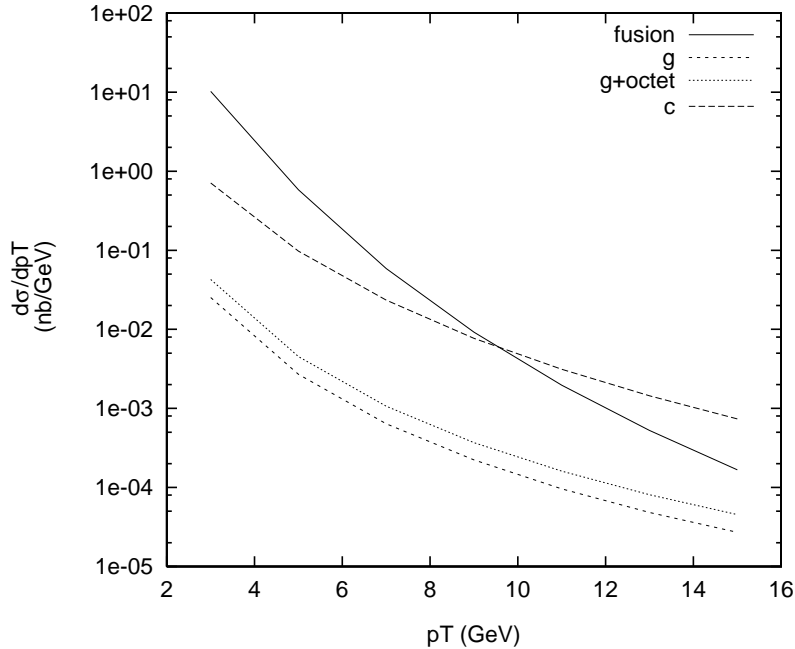,
             bbllx=30pt,bblly=160pt,bburx=540pt,bbury=660pt,
             width=11cm,clip=}

\vspace*{-1.75cm}
\begin{center}
\parbox{12cm}{
\caption[dummy]{\label{fig_grs} \small Transverse momentum distribution 
  $\mbox{d}\sigma/\mbox{d}p_\perp$ for $J/\psi$ photoproduction at the
  photon-proton centre of mass energy $\nu = 100$~GeV. The solid line
  represents the (leading-order) fusion contribution and the
  dashed-dotted line the charm quark fragmentation contribution.  The
  dotted and dashed lines represent the gluon fragmentation
  contributions with and without a color-octet component for the $S$
  state. The cut on the inelasticity parameter $z$ is $0.1 < z <
  0.9$. From Ref.\cite{GRS95}.}
}
\end{center}

\end{figure}

A quantitative analysis of the fragmentation contributions to $\jpsi$
photoproduction at HERA has been performed in ref.\cite{GRS95} (see
also ref.\cite{SA94} for an earlier analysis within the color-singlet
model). The authors have considered the fragmentation of gluons and
charm quarks produced at leading order via the Compton- and
Bethe-Heitler processes, $\gamma + q \to q + g$ and $\gamma + g \to c
+ \bar{c}$, respectively. It appears that at large transverse momenta,
$p_T\;\simgt\;10$~GeV, color-singlet charm quark fragmentation
dominates over the photon-gluon fusion process while
gluon-fragmentation is suppressed by an order of magnitude over the
whole range of $p_T$, see fig.\ref{fig_grs} \cite{GRS95}.  Color-octet
contributions to the gluon fragmentation function considerably enhance
the large-$z$ region, but have no strong effect in the inelastic
domain $z\;\simlt\; 0.9$.  Thus, the information that will be obtained
from large $p_T$ production of $\jpsi$ particles at HERA can be used
to study the charm quark fragmentation mechanism and is complementary
to the analyses performed at the Tevatron where the large $p_T$ region
is dominated by gluon fragmentation.  However, since the cross section
for $\jpsi$ photoproduction at $p_T\;\simgt\;10$~GeV is at the most
${\cal O}(1 pb)$, a large luminosity is required to probe
fragmentation contributions to quarkonium production at HERA.

\section{Resolved photon contributions}

The photoproduction of a $\jpsi$ particle can also take place via a so
called resolved photon interaction, where the photon couples through
one of its hadronic components. For $\jpsi$ production within the 
color-singlet model the process goes like:
\beq\label{psires}
\gamma p \to g_\gamma + g_p \to c\bar c[^3S_1,\underline{1}] + g 
\to \jpsi + g
\eeq
meaning that the gluon coming from the photon fuses with the one
coming from the proton to give a $^3S_1$ color-singlet $c\bar c$ pair
which subsequently hadronizes into a $\jpsi$. This process has been
extensively analyzed in the past (see for instance \cite{JST92}) and
found to contribute to the overall cross section only marginally
everywhere but in the low-$z$ region.  More contributions, coming from
production and radiative decay of $\chi$ states, are also expected to
be present. These terms also probe the quark content of the photon.
They are expected to be of comparable size with the $\jpsi$ production
process (\ref{psires}) described above (see \cite{JST92} for a survey
of these and other resolved production mechanisms).

Within the factorization approach, however, more channels have to be
considered, where the $\jpsi$ production goes via color-octet $c\bar
c$ pairs.  Namely, the following processes contribute:
\beqa
&&g_\gamma + g_p \to c\bar c[^1S_0,\underline{8}] + g\nonumber \\
&&g_\gamma + g_p \to c\bar c[^3S_1,\underline{8}] + g\nonumber \\
&&g_\gamma + g_p \to c\bar c[^3P_J,\underline{8}] + g\nonumber 
\,.
\eeqa
They are in every respect analogous to the ones which have been argued
to greatly increase the $\jpsi$ production in $p\bar p$ collisions at
the Tevatron. They must therefore be carefully considered here to
understand whether or not they change the picture established by the
CSM.

Using the matrix elements squared evaluated in ref.\cite{CL96} and the
choice of nonperturbative parameters displayed in Table \ref{table1}
we can calculate the cross sections and compare them with the CSM
ones. The results are shown in Table \ref{table2}, with a minimum
$p_T$ cut needed to screen the collinear singularity present in the
$^1S_0$ and $^3P_J$ channels. 

\begin{table}[htb]
\begin{center}
\begin{tabular}{|c|c|r|} 
\hline
\multicolumn{2}{|c|}{}        & $\sigma_{\gamma p}$ (nb) \\
\multicolumn{2}{|c|}{Channel} & $\sqrt{s}$ = 100 GeV  \\
\multicolumn{2}{|c|}{}        & {$p_T>1$ GeV}   \\
\hline
Direct & $^3S_1,\underline{1}$ & 13.68  \\
\hline
            & $^3S_1,\underline{1}$ & .48    \\
            & $^1S_0,\underline{8}$ & .79  \\
            & $^3S_1,\underline{8}$ & .34  \\
Resolved    & $^3P_0,\underline{8}$ & 1.61  \\
            & $^3P_1,\underline{8}$ & .50 \\
            & $^3P_2,\underline{8}$ & 2.06   \\
\hline
\end{tabular}
\parbox{12cm}{
\caption{
\label{table2}
\small Results for the total cross sections (in nb).}
}
\end{center}
\end{table}

\begin{figure}[t]
\begin{center}
\epsfig{file=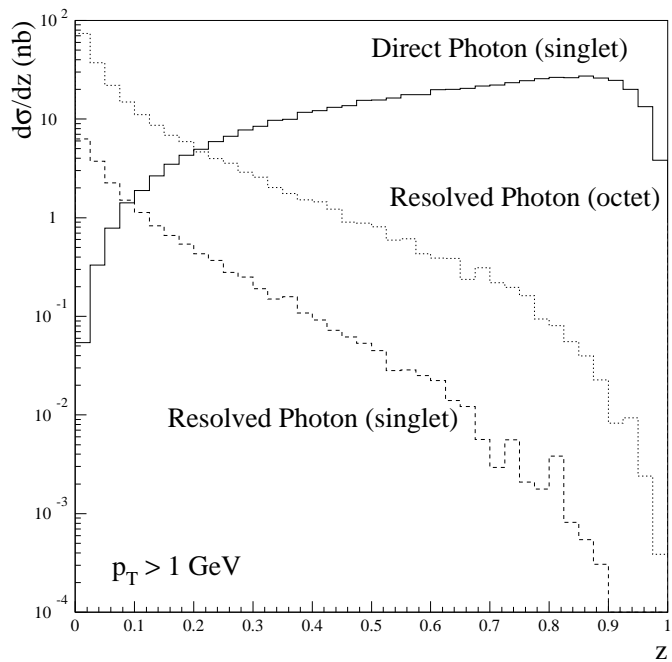,
              bbllx=30pt,bblly=160pt,bburx=540pt,bbury=660pt,
            width=9.cm,clip=}
\parbox{12cm}{
\caption{\label{fig1}\small Comparison of the inelasticity distributions for
direct (singlet) and resolved (singlet and octet) $\jpsi$ photoproduction.
}
}
\end{center}
\end{figure}

These results make clear that the color-octet channels could provide a
non-negligible increment of the overall $\jpsi$ photoproduction cross
section.  It's therefore worth studying in more detail which regions
of phase space will be mostly affected. A close look to the
differential distributions of experimental interest shows that the
behaviour of the color-octet contributions to the resolved channels is
similar to that of the color-singlet one. Namely, it is only visible
in the low-$z$ region. Fig.\ref{fig1} shows the $z$ distribution for
the direct photon color-singlet channel (full line) compared with the
old CSM resolved photon contribution (dashed line) and the new
resolved photon color-octet one (dotted line). The resolved
contributions can be seen to be enhanced by the color-octet terms, but
the qualitative picture of them being visible only in the low-$z$
region remains unchanged.

\section{$\chi$ photoproduction}

For the production of $S$-wave quarkonia, like $\jpsi$, $\psi'$ or
$\Upsilon$, the factorization approach coincides with the
color-singlet model in the nonrelativistic limit $v\to 0$. In the case
of $\chi$ bound states, however, color-singlet and color-octet
mechanisms contribute at the same order in $v$ to annihilation rates
and production cross sections and must therefore both be included for
a consistent calculation \cite{BBL92}.  The production of $\chi$
states plays a peculiar role in photon-hadron collisions. Unlikely
what happens in hadron-hadron collisions, were $\chi$s are copiously
produced by $gg$, $gq$ and $q\bar q$ interactions, the process
\beq
\gamma + g \to c\bar c[^3P_J, \underline{1}] + g
\eeq
namely the production of a color-singlet $c\bar c$ pair with the
$\chi$ quantum numbers, happens to be forbidden at leading order in
$\as$. This is due to the color factor of the two gluons having to be
symmetric in order to produce a color-singlet states. The gluons could
therefore be replaced by two colorless photons, and the process $\chi
\to \gamma\gamma\gamma$ is know to be forbidden by the request of
charge conjugation invariance.

To have $\chi$ production initiated by a direct photon we must
therefore either go to higher orders within the CSM, adding one gluon
to the leading order diagram, or consider color-octet mediated
channels.

Within the factorization approach, indeed, $\chi$ production can still
take place at the leading ${\cal O}(\alpha \alpha_s^2)$, provided it
is a color-octet $^3S_1$ charm pair which is produced in the hard
interaction and subsequently hadronizes into a physical $\chi$
particle. The following two processes contribute:
\beqa
&&\gamma + g \to c\bar c[^3S_1,\underline{8}] + g\nonumber \\
&&\gamma + q(\bar q) \to c\bar c[^3S_1,\underline{8}] + q(\bar q)\nonumber 
\eeqa
The first of these two reaction is by far the dominant one in the HERA
energy range and at low transverse momentum. Within the spirit of the
factorization approach, the cross section is given by the same short
distance cross section evaluated for color-octet $\jpsi$
photoproduction (see eq.(\ref{eq:bbl}) and the second process of
eq.(\ref{jpsioctet})), times the appropriate matrix elements $\langle
{\cal O}^{\chi_J}[^3S_1,\underline{8}]\rangle$. These matrix elements
have also been fitted to the Tevatron data in \cite{TEV1,CL96}, and
found to be of order $10^{-2}$ GeV$^3$. Using, for the sake of
simplicity, $\langle {\cal O}^{\chi_0}[^3S_1,\underline{8}]\rangle =
10^{-2}$ GeV$^3$ and the relation
\beq
\langle {\cal O}^{\chi_J}[^3S_1,\underline{8}]\rangle = (2J+1) 
\langle {\cal O}^{\chi_0}[^3S_1,\underline{8}]\rangle
\eeq
we find
\beq
\sum_{J}\sigma(\gamma p \to \chi_J) = 
9 \langle {\cal O}^{\chi_0}[^3S_1,\underline{8}]\rangle \times 0.9 
\mathrm{~nb~GeV}^{-3} \simeq 80 \mathrm{~pb}
\eeq
at $\sqrt{s}$ = 100 GeV  and with a minimum $p_T$ cut of 3 GeV.

It is worth noticing here a pretty large discrepancy with the result
of ref.\cite{MA}. Despite using a nonperturbative matrix element about
a factor of two smaller than ours, it finds $\sigma(\gamma p \to
\chi_1) = 0.13~\mathrm{nb}$, which is about a factor of five larger
than our result.

\section{Associated $\jpsi + \gamma$ production}

A particularly distinctive experimental probe of the relevance of
color-octet contributions in quarkonia production can be the
observation of the exclusive process given by the associated
production of a $\jpsi$ and a photon \cite{cgk}\footnote{see also
  ref.\cite{KR93} for an analysis within the color-evaporation
  model.}:
\beq
\gamma p \to \jpsi + \gamma 
\eeq
Within the CSM this process can only undergo via the resolved photon
channel, since at least two gluons must couple to the heavy quark line
to produce a color-singlet $c\bar c$ pair:
\beq
\gamma p \to g_\gamma + g_p \to c\bar c[^3S_1,\underline{1}] + \gamma 
\to \jpsi + \gamma 
\eeq
A big suppression of the cross section, and the characteristic $z$
distribution typical of resolved photon processes, peaked at low $z$,
can therefore be expected.

\begin{figure}[htb]
\begin{center}
\epsfig{file=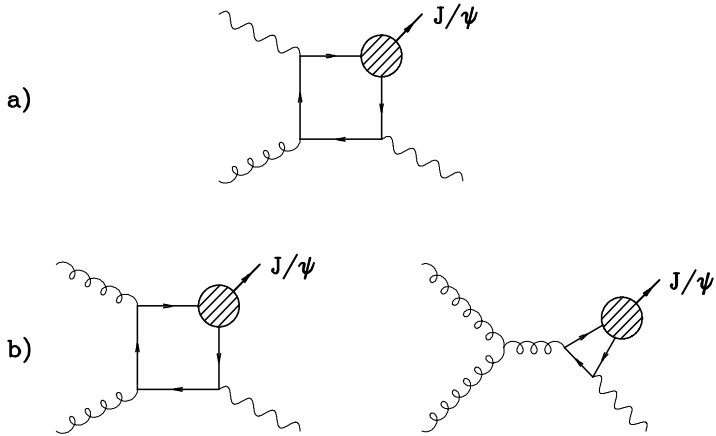,width=9.5cm}
\parbox{12cm}{
\caption{\label{fig2}\small Diagrams contributing within the factorization
approach to $\jpsi$ + photon associated production in direct (a)
and resolved (b) photon collisions. In (b) the right diagram contributes 
to $^1S_0$ and $^3P_J$ states production only.}
}
\end{center}
\end{figure}

Within the factorization approach, on the other hand, other $c\bar c$
states can contribute and, in particular, the reaction can now proceed
also via a direct photon channel.  Indeed, the following processes are
now possible:

\vspace{.5cm}
\noindent
Direct photon (fig.\ref{fig2}a):
\beqa
&&\gamma + g_p \to c\bar c[^1S_0,\underline{8}] + \gamma\nonumber \\
&&\gamma + g_p \to c\bar c[^3S_1,\underline{8}] + \gamma\nonumber \\
&&\gamma + g_p \to c\bar c[^3P_J,\underline{8}] + \gamma\nonumber
\eeqa

\noindent
Resolved photon (fig.\ref{fig2}b):
\beqa
&&g_\gamma + g_p \to c\bar c[^1S_0,\underline{8}] + \gamma\nonumber \\
&&g_\gamma + g_p \to c\bar c[^3S_1,\underline{1}] + \gamma\qquad(\mathrm{
Standard~CSM~process})\nonumber\\
&&g_\gamma + g_p \to c\bar c[^3S_1,\underline{8}] + \gamma\nonumber \\
&&g_\gamma + g_p \to c\bar c[^3P_J,\underline{8}] + \gamma\nonumber 
\eeqa

Though the nonperturbative matrix elements which mediate the
hadronization of the color-octet $c\bar c$ pairs to a $\jpsi$ are
suppressed with respect to the color-singlet one (see Table
\ref{table1}), we can however still expect the two following features:
\begin{itemize}
\item[i)] the production of color-octet states via a direct process -
  rather than a resolved one - will at least partially compensate for
  the smaller matrix elements. We should therefore expect an increase
  in the overall cross section;
\item[ii)] the $z$ distribution of the $\jpsi$, will be more peaked
  towards one. This is again due to the presence of a direct photon
  coupling as opposed to the resolved one, where the $g_\gamma$ only
  carries part of the photon energy into the reaction.
\end{itemize}
These (and especially the second one) are the reasons why we expect
this process to be a sensitive probe to color-octet components.

\begin{table}[htb]
\begin{center}
\begin{tabular}{|c|c|r|} 
\hline
\multicolumn{2}{|c|}{} & $\sigma_{\gamma p} (pb)$ \\
\multicolumn{2}{|c|}{Channel} & {$\sqrt{s}$ = 100 GeV} \\
\multicolumn{2}{|c|}{} & {$p_T>1$ GeV}   \\
\hline
       & $^1S_0,\underline{8}$  &   --  \\
Direct & $^3S_1,\underline{8}$  & 7.67  \\
       & $^3P_J,\underline{8}$  &   --  \\
\hline
            & $^1S_0,\underline{8}$ &   .35  \\
            & $^3S_1,\underline{1}$ & 16.70   \\
            & $^3S_1,\underline{8}$ &   .27  \\
Resolved    & $^3P_0,\underline{8}$ &  1.03  \\
            & $^3P_1,\underline{8}$ &   .14  \\
            & $^3P_2,\underline{8}$ &   .97  \\

\hline
\end{tabular}
\parbox{12cm}{
\caption{
\label{table3}
\small Results for the total cross sections of $\jpsi + \gamma$
photoproduction (in pb). At leading-order, the ${}^1S_0$ and ${}^3P_J$
direct contributions vanish identically. } }
\end{center}
\end{table}

The total cross sections (with a minimum $p_T$ cut of 1 GeV) are shown
in Table \ref{table3} for $\gamma p$ collisions at a cm energy of 100
GeV (see ref. \cite{cgk} for more details).  The bulk of the cross
section can be seen to come from the standard CSM channel $gg \to
c\bar c[^3S_1,\underline{1}] + \gamma$. However, the direct channel
gives a non negligible contribution, amounting to about 25\% of the
overall cross section of $\sim$ 27 pb. This increase is however far
too small to be reliably used to assess the presence of the
color-octet terms, given the smallness of this cross section (in the
picobarn region) and the large theoretical uncertainties involved
(like the charm mass, the strong coupling and the nonperturbative
matrix elements values).

On the other hand, the study of the differential distributions can 
make easier to disentangle the color-octet contributions from the
standard color-singlet one.

\begin{figure}[t]
\begin{center}
\epsfig{file=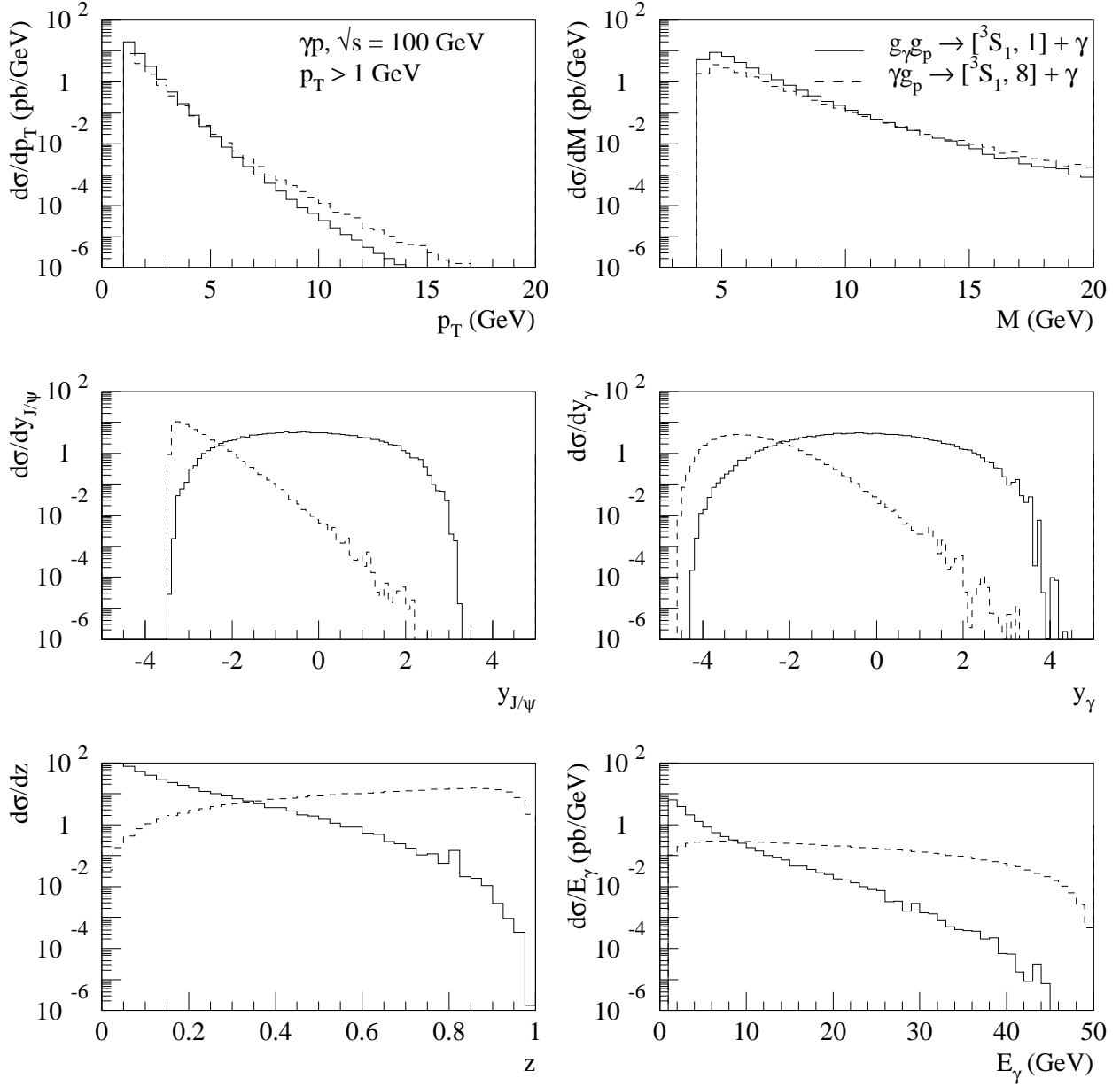,
              bbllx=30pt,bblly=160pt,bburx=540pt,bbury=660pt,
             width=17.cm,clip=}
\parbox{12cm}{
\caption{\label{fig3}\small Differential distributions in $\gamma p$ collision
at $\protect\sqrt{s} = 100$ GeV. A minimum $p_T$ cut of 1 GeV is applied.
}
}
\end{center}
\end{figure}

We therefore show in fig.\ref{fig3} the differential distributions
related to the total $\gamma p$ cross sections at $\sqrt{s} = $ 100
GeV with a minimum $p_T$ cut of 1 GeV, presented in Table
\ref{table3}. The distributions due to [$^3S_1$,\underline{1}]
production in resolved photon collision (continuous line) and to
[$^3S_1$,\underline{8}] production in direct photon collision (dashed
line) only are shown. The distributions due to the other color-octet
processes do indeed present the same features as the ones of
[$^3S_1$,\underline{1}], being also produced in resolved photon
interactions, but are suppressed in magnitude, as can be seen from
Table \ref{table3}. Their inclusion would therefore not change the
picture we are going to discuss.

Fig.\ref{fig3} compares the result of the CSM with that due to the
production of a color-octet $^3S_1$ state in direct photon collision,
fig.\ref{fig2}a, as predicted by the factorization approach. As
expected, the effect of the direct photon coupling can easily be seen
in at least some of the plots. While the $p_T$ of the $\jpsi$ and the
invariant mass distribution of the $\jpsi$-$\gamma$ pair are pretty
similar for the two processes, the $z$, rapidity and photon energy
distributions do indeed show a strikingly different behaviour.

Recalling that we put ourselves in the so-called ``HERA-frame'', with
the photon (or the electron) traveling in the direction of negative
rapidities, we notice how the direct photon coupling favours the
production of the quarkonium and of the photon in the negative
rapidities region. This contrasts the case of resolved photon
production of color-singlet $^3S_1$ states, which are uniformly
produced around the central region.

As for the $z$ distribution, the resolved photon process predicts a
decrease of the cross section going towards the high-$z$ region. The
direct photon process does on the other hand predict the opposite
behaviour: the cross section now increases going towards $z=1$. The
small dip in the last few bins is due to the minimum $p_T$ cut.

Similarly to the $z$ distribution behaves the photon energy one, which
is predicted to be much harder in direct photon processes.

These distributions (which were also checked to be robust with respect
to a higher $p_T$ cut, so as to be sure of the absence of $p_T^{min}$
effects) could already be good experimental discriminators:
observation of a substantial fraction of $\jpsi + \gamma$ events in
the high $z$ region would provide good evidence for the presence of
color-octet contributions to the overall cross section.

\section{$\jpsi$ production in deep inelastic scattering}\label{DIS}

Leptoproduction of quarkonium states has been extensively studied in the 
framework of the color-singlet model \cite{CS}. Though the color-singlet 
contribution can explain inelastic leptoproduction of $\jpsi$ it can not 
explain the total cross section. Thus, in order to arrive at a complete 
description of $\jpsi$ leptoproduction, the color-octet contributions to 
the $\jpsi$ production rate have been calculated in ref.\cite{FMM}. 
The authors obtain the following expression for the
differential subprocess cross section:
\begin{eqnarray}
Q^2 {d \hat{\sigma} \over d Q^2} (eg \to e \jpsi) & = &
{2 \pi^2 e^2_c \alpha_s \alpha^2 \over m_c \hat{s}} \; 
\int {dy \over y^2} \Bigg\{ {1+(1-y)^2 \over  y } 
\nonumber \\
& & \times \left[ y \langle  {\cal O}^\jpsi [^1S_0,\underline{8}] \rangle + 
{\langle  {\cal O}^\jpsi [^3P_0,\underline{8}] \rangle \over m^2_c} 
{3Q^2+7(2m_c)^2 \over \hat{s}} \right]  
\nonumber \\
& & - {\langle {\cal O}^\jpsi [^3P_0,\underline{8}] \rangle \over m^2_c}
{8(2m_c)^2 Q^2 \over \hat{s}^3} 
 \Bigg\}
\delta \Big( \hat{s}y - (4m^2_c + Q^2) \Big) \; ,
\label{fullsubcs}
\end{eqnarray}
where $\sqrt{\hat{s}}$ is the subprocess center-of-mass energy, $Q^2$
is the negative invariant mass of the photon, and $y$ is the momentum
fraction of the $\jpsi$ relative to the incoming electron.  To obtain
the cross section one convolutes the expression given in
eq.(\ref{fullsubcs}) with the gluon distribution function. The
hadronization of the $c\bar{c}$ pair, produced initially in a
color-octet state with spin/angular-momentum quantum numbers
$^{2S+1}L_J$, into a $\jpsi$ bound state is parametrized by the NRQCD
matrix elements $\langle {\cal O}^\jpsi [^1S_0,\underline{8}] \rangle
$ and $\langle {\cal O}^\jpsi [^3P_0,\underline{8}] \rangle $.  Note
that it is precisely the matrix elements appearing in
eq.(\ref{fullsubcs}) that are at the heart of the discrepancy between
the CDF measurement and the photoproduction results.

There is an important point regarding the differential cross section
presented in eq.(\ref{fullsubcs}). In principle this result is valid
for all values of $Q^2$, however, there are corrections due to higher
twist terms that have been neglected in the derivation of the factored
form of the cross section. These higher twist terms are suppressed by
powers of $m^2_c / Q^2$, and will, therefore, vanish for {}$Q^2 \gg
m^2_c$. Thus it is necessary to compare the theoretical results
presented in eq.(\ref{fullsubcs}) to experimental data in the large
$Q^2$ regime. In the limit where $\hat{s}, Q^2 \gg m^2_c$,
eq.(\ref{fullsubcs}) reduces to
\begin{eqnarray}
\lim_{\hat{s}, Q^2 \gg m^2_c} 
Q^2 {d \hat{\sigma} \over d Q^2} (eg \to e \jpsi) & \longrightarrow  &  
{2 \pi^2 e^2_c \alpha_s \alpha^2 \over m_c \hat{s}} \;
\int dy  {1+(1-y)^2 \over  y^2 }
\nonumber \\
& & \times  \Bigg\{ \langle {\cal O}^\jpsi [^1S_0,\underline{8}] \rangle + 3 
{\langle  {\cal O}^\jpsi [^3P_0,\underline{8}] \rangle \over m^2_c } \Bigg\} 
\; \delta\big( y - {Q^2 \over \hat{s}} \big) \; .
\nonumber \\
& & \mbox{}
\label{limitsubcs}
\end{eqnarray}
Note that for large $Q^2$ the linear combination {}$\langle {\cal
  O}^\jpsi [^1S_0,\underline{8}] \rangle + 
  3 \langle {\cal O}^\jpsi [^3P_0,\underline{8}] \rangle /
m^2_c$ is determined. This is precisely the linear combination of
NRQCD matrix elements that is measured at CDF. Therefore $\jpsi$
leptoproduction can provide an independent means of determining this
linear combination of NRQCD matrix elements in a manner different from
the CDF measurement.

Currently there exists experimental data on the production of $\jpsi$
in $\mu N$ collisions~\cite{emc}, however, the values of $Q^2$ probed in
this experiment are too low to be in the asymptotic region, and the
error on the experimental measurements are too large to allow for an
accurate determination of the color-octet matrix elements. The 
high-statistics measurements to be expected in the future at 
\mbox{HERA} could definitely help to improve the situation and to 
constrain the color-octet matrix elements. 

\section{Conclusion}
We have worked out and reviewed the impact of color-octet
contributions on quarkonium production in photon-proton collisions and
deep-inelastic scattering at HERA. Photoproduction of $\jpsi$,
$\psi'$, $\Upsilon$ and $\chi$ states, associated $\jpsi + \gamma$
production, fragmentation and resolved-photon contributions as well as
deep inelastic $\jpsi$ production have been discussed. We have shown
how these reactions can be used to constrain the color-octet matrix
elements and test the picture of quarkonium production as developed in
the context of the NRQCD factorization approach.

\vspace*{5mm}

\noindent
{\bf Acknowledgements}

\noindent Numerous and fruitful conversations with Mario Greco are
gratefully acknowledged. Section~\ref{DIS} was prepared in
collaboration with Sean Fleming. We are grateful to him and to Rohini
Godbole and K.~Sridhar for providing Figure \ref{fig_grs} at short
notice.


\end{document}